# Manifestation of Spin Selection Rules on the Quantum Tunneling of Magnetization in a Single Molecule Magnet

J. J. Henderson[1], C. Koo[2], P. L. Feng[3], E. del Barco[1,*], S. Hill[4], I. S. Tupitsyn[5], P. C. E. Stamp[5] and D. N. Hendrickson[3]

[1]*Department of Physics, University of Central Florida, Orlando, FL 32816, USA*

[2]*Department of Physics, University of Florida, Gainesville, FL32611, USA*

[3]*Department of Chemistry and Biochemistry, University of California at San Diego, La Jolla, CA 92093, USA*

[4]*National High Magnetic Field Laboratory and Department of Physics, Florida State University, Tallahassee, FL 32310, USA*

[5]*Pacific Institute for Theoretical Physics, University of British Columbia, 6224 Agricultural Road, Vancouver, B.C., Canada, V6T 1Z1*

[*]*Electronic mail: delbarco@physics.ucf.edu*

ABSTRACT:

We present low temperature magnetometry measurements on a new $Mn_3$ single-molecule magnet (SMM) in which the quantum tunneling of magnetization (QTM) displays clear evidence for quantum mechanical selection rules. A QTM resonance appearing only at elevated temperatures demonstrates tunneling between excited states with spin projections differing by a multiple of three: this is dictated by the $C_3$ symmetry of the molecule, which forbids pure tunneling from the lowest metastable state. Resonances forbidden by the molecular symmetry are explained by correctly orienting the Jahn-Teller axes of the individual manganese ions, and by including transverse dipolar fields. These factors are likely to be important for QTM in *all* SMMs.

The fascinating observations of basic quantum phenomena offered by single-molecule magnets (SMMs) [1-12] and their potential use in quantum information technologies [2] has boosted the interest in these materials, and resulted in a better understanding of the basic laws governing the quantum dynamics of magnetization in nanoscale systems. In the bottom-up approach provided by supramolecular chemistry, the quantum mechanical properties of SMMs are essentially dictated by the molecular composition and configuration. Molecular and crystallographic symmetries thus play an essential role in the quantum tunneling of the magnetization (QTM) in SMMs. An elegant example of this is the observation of Berry phase quantum interference effects between equivalent tunneling trajectories, which are determined by transverse anisotropy terms in the Hamiltonian directly imposed by the molecular symmetry [6,9].

Crucially, symmetry also enforces spin selection rules in tunneling, only allowing tunneling transitions with a change of spin projection, $|\Delta m_S|$, equal to the order of the creation/annihilation spin operators associated with the transverse anisotropy terms in the Hamiltonian. Thus, e.g., molecules with $S_4$ symmetry (e.g. $Mn_{12}$tBuAc; tetragonal crystal lattice), should only tunnel when $|\Delta m_S| = 4n$, where $n$ is an integer; and molecules with $C_3$ (e.g. $Mn_3$ [13-19]; trigonal lattice) or $D_2$ (e.g. $Fe_8$; triclinic lattice) symmetries should only tunnel when $|\Delta m_S| = 3n$ and $2n$, respectively. Surprisingly, in all experimental studies on SMMs to date, QTM is found to occur at all resonances, regardless of the spin selection rules imposed by the molecular symmetry, of which only subtle manifestations have been reported so far [6,20,21]. There are many possible reasons for this. One is that intrinsic disorder (coming from solvent loss, ligand disorder or crystalline defects) lowers the local molecular symmetry so that an applied longitudinal field induces a distribution of transverse magnetic field components [7-9]. Other possibilities include the dynamic fields from nuclear spins, and dipolar interactions [22]. What has been lacking so far is a clear-cut experimental demonstration of the role of molecular symmetry in QTM, in which all of these extraneous effects are either eliminated or accounted for.

In this Letter we present such a demonstration, using a high-quality crystal of a simple $Mn_3$ complex. The crystalline quality is demonstrated by the unsurpassed sharpness of the X-ray diffraction and EPR absorption peaks [13,14]. We observe a sequence of very sharp QTM steps, out of which one resonance appears only at high temperature. We show that the high temperature relaxation at this resonance is associated with tunnel transitions involving excited states with $|\Delta m_S| = 3n$, as dictated by the $C_3$ symmetry of the molecule, which forbids tunneling from the lowest metastable state at low temperature. In addition, a rotation of the local zero-field-splitting (ZFS) tensors following the tilts of the Jahn-Teller axes of the $Mn^{III}$ ions, in combination with intermolecular dipolar interactions, accounts for the observed behavior in all QTM resonances, including transitions forbidden by the molecular symmetry.

We have studied the two complexes [NEt$_4$]$_3$[**Mn$_3$**Zn$_2$(salox)$_3$O(N$_3$)$_6$**Cl$_2$**] and [NEt$_4$]$_3$[**Mn$_3$**Zn$_2$(salox)$_3$O(N$_3$)$_6$**Br$_2$**], henceforth Mn$_3$-Cl and Mn$_3$-Br, respectively [13,14]. The metallic cores of these complexes are comprised of a $\mu_3$-oxo-centered triangle of $Mn^{3+}$ ions and two capping $Zn^{2+}$ ions located above and below the Mn$_3$ plane, resulting in a rigid trigonal bipyramidal structure. The diamagnetic $Zn^{2+}$ ions and bulky [NEt$_4$]$^+$ cations isolate the Mn$_3$ magnetic core from intermolecular magnetic interactions, as evident from the absence of significant intermolecular contacts and the 10.30 Å minimum separation between $Mn^{III}$ ions in neighboring molecules. Both complexes crystallize in the trigonal space group $R3c$ as racemic mixtures of $C_3$-symmetric chiral molecules (with equal population of molecules with opposite chirality, rotated by 27 degrees about the $C_3$ axis with respect to each other). Neither structure contains solvate molecules, which is quite rare for SMMs and likely explains the extremely high resolution spectroscopic data (solvents evaporate easily, causing disorder [23]). Ferromagnetic exchange interactions between $Mn^{3+}$ ions are propagated by the central $\mu_3$-oxo ion and through the coordinating oxime, resulting in a molecular spin $S = 6$ ground state at low temperature. These structural and crystallographic properties differentiate $Mn_3$-Cl and $Mn_3$-Br from other ferromagnetic $Mn_3$ triangles, each of which possesses appreciable intermolecular interactions, low molecular symmetry, or co-crystallized solvate molecules [15-19]. We observe only subtle differences in the magnetic behavior of $Mn_3$-Cl and $Mn_3$-Br; we will thus consider them identical in the context of any discussion in this Letter.

Magnetization hysteresis measurements were carried out on sub-millimeter size single crystals of both $Mn_3$-Cl and $Mn_3$-Br, using a high-sensitivity micro-Hall effect magnetometer [24] in the temperature range of 0.3-2.6 K. Figure 1 shows the first derivative of the magnetization (for $Mn_3$-Cl) plotted versus the longitudinal magnetic field, $H_L$ (along the easy anisotropy axes of the molecules), at different temperatures. Narrow peaks corresponding to the $k = 0, 1, 2$ and $3$ QTM resonances are clearly observed at almost regular field intervals ($\Delta H \sim 0.85$ T). Resonance $k = 1$ (0.85 T) is not visible at the lowest temperatures in Fig. 1; it appears only upon application of a transverse field (see Fig. 2), or upon raising the temperature above 1.5 K, when it appears suddenly at a lower field value (0.80 T). To the best of our knowledge, this is the first occasion in which a QTM resonance (e.g. $k = 1$) within a series of resonances is found to be absent [21], while lower and higher resonances are observed. As we show below, this constitutes definite evidence for the influence of spin selection rules for QTM in a SMM.

The observed shift of all resonances to lower fields with increasing temperature indicates a transition from the pure quantum tunneling regime, in which the relaxation occurs from the ground spin state, to thermally activated tunneling between excited states [4]. The fact that the resonances associated with excited states appear at lower field values is indicative of a fourth order uniaxial anisotropy term in the Mn$_3$ Hamiltonian (i.e. $B_4^0 \hat{O}_4^0$), coming from a relatively weak exchange interaction constant ($J$) between the manganese ions (comparable to the single-ion ZFS interaction) in the multi-spin Hamiltonian given by [25]:

$$H = \sum_i \left( \vec{s}_i \cdot \vec{R}_i^T \cdot \vec{D} \cdot \vec{R}_i \cdot \vec{s}_i - \mu_B \vec{s}_i \cdot \vec{g} \cdot \vec{B} \right) + \tfrac{1}{2} \sum_{i,j(i \neq j)} \vec{s}_i \cdot \vec{J} \cdot \vec{s}_j . \qquad (1)$$

Here the first term represents the local magnetic anisotropy of the $i$-th ion, $\vec{D}$ being the ZFS (diagonal) tensor given by $D_{xx} = e$, $D_{yy} = -e$ and $D_{zz} = -d$, with $d$ and $e$ the uniaxial and second order transverse anisotropy parameters, respectively. $\vec{R}_i$ is the Euler matrix specifying the anisotropy axes of the three Mn ions in the molecule, defined by the Euler rotation angles $\alpha_i$, $\beta_i$ and $\gamma_i$. The second term is the Zeeman coupling to the applied magnetic field, and the last term is the exchange interaction between neighboring ions. The positions of the QTM resonances observed in both the quantum and thermally activated regimes (Figs. 1-3) can accurately be accounted for using the following set of parameters: $s_i = 2$, $d = 4.2$ K, $e \sim 0.9$ K, isotropic $g = 2$ and $J = -4.88$ K. The single-ion second order anisotropy ($e$) is needed to explain the observed QTM rates. Similar $d$ and $e$ values have been reported for Mn$^{III}$ ions in the literature (see, for example, Ref. [26]). We additionally rotated the anisotropy axes for the three Mn$^{III}$ ions such that $\alpha_i = 8.5^o$ (with $\gamma_i = 0$), and $\beta_1 = 0$, $\beta_2 = 120^o$ and $\beta_3 = 240^o$, in order to account for the local tilts of the Jahn-Teller axes and to preserve the $C_3$ rotation symmetry [13,14].

Fig. 2 shows the magnetization curve obtained at 300 mK for the Mn$_3$-Cl single crystal shown in Fig. 1. From the measured changes in magnetization at the resonances, we can obtain a rough estimate of the tunnel splittings associated with the superposition of different spin states at each resonance. At resonance $k = 1$, the splitting ($\Delta_{-6,+5} \sim 1 \times 10^{-6}$ K) is found to be one order of magnitude smaller than those at resonances $k = 0$ ($\Delta_{-6,+6} \sim 7 \times 10^{-6}$ K), $k = 2$ ($\Delta_{-6,+4} \sim 1 \times 10^{-5}$ K) and $k = 3$ ($\Delta_{-6,+3} > 1 \times 10^{-5}$ K) [27]. Interestingly, the absent resonance ($k = 1$) can clearly be observed by decreasing the field sweep rate (not shown) or by applying a transverse field. The insets to Fig. 2 show the growth of the $k = 1$ peak (lower-right) and the calculated QTM probability [28] at resonances $k = 1$ and $k = 2$ (upper-left) as a function of the transverse field magnitude. The curvature of the probability at $k = 1$ near zero magnetic field is indicative of a saturation caused by intermolecular dipolar interactions, whose magnitude (~250 G) can be estimated from the width of the peaks.

The $C_3$ symmetry of the Mn$_3$ complexes should only allow tunneling between states on opposite sides of the barrier with spin-projection ($m_S$) values differing by a multiple of three. According to this selection rule, the only resonances observable below the crossover temperature separating the pure quantum tunneling regime from the thermally activated one should be $k = 0$ (tunneling from $m_S = -6$ to $m_S = +6$, $\Delta m_S = 12 = 4 \times 3$) and $k = 3$ (tunneling from $m_S = -6$ to $m_S = +3$, $\Delta m_S = 9 = 3 \times 3$). Fig. 3 shows the energy of the $m_S$ levels as a function of the longitudinal field, calculated from exact diagonalization of the Hamiltonian in Eqn. (1). The first allowed QTM transition for each resonance is indicated at the anti-crossings between the respective spin levels (blue squares). Note that the lowest levels involving the $k = 1$ and $k = 2$ resonances cross exactly and, therefore, tunneling is forbidden (red circles in Fig. 3) since this would involve transitions that are not a multiple of three (i.e. $\Delta m_S = 11$ for ground state at $k = 1$ and $\Delta m_S = 10$ and 8 for the ground and first excited states at $k = 2$). This explains why resonance $k = 1$ only becomes visible at high temperatures in Fig. 1, since an excited state needs to be populated for the QTM to take place (i.e. the transition from $m_S = -5$ to $m_S = +4$, $\Delta m_S = 9 = 3 \times 3$). Consequently, the temperature dependence of resonance $k = 1$ constitutes firm evidence for the spin selection rules imposed on QTM in a SMM.

Following the same arguments, and contrary to the experimental observations, one should expect resonance $k = 2$ to be absent at low temperatures as well (see Fig. 3). This inconsistency can be understood in terms of tilting of the Jahn-Teller axes of the manganese ions within the molecule. This can clearly be observed in Fig. 4, which shows the dependence of the ground state tunnel splittings on the magnitude of the transverse field for all resonances observed in the experiment ($k = 0$-3), calculated via exact diagonalization of the Hamiltonian in Eqn. (1) using the parameters given above [29] for two different situations: a) With the Jahn-Teller axes aligned with the $z$-axis, i.e. $\alpha = 0$ (thin lines in Fig. 4); and, b) including a tilting of the Jahn-Teller axes of each ion away from the crystallographic $z$-axis (molecular easy-axis), given by the Euler angle $\alpha = 8.5^\circ$ (thick lines in Fig. 4), determined from X-ray crystallography data [13,14]. The tilting is represented by a sketch in Fig. 4. In both cases, the tunnel splittings at resonances $k = 1$ and $k = 2$ completely vanish in the absence of a transverse field ($\Delta_{k=1} = \Delta_{k=2} = 0$) [30]. If the Jahn-Teller axes are not tilted, large transverse fields ($H_T > 0.2$ T) are required to bring the tunnel splittings up to the magnitudes observed in the experiment. However, the inclusion of a tilting of the Jahn-Teller axes by 8.5º has a profound influence on the transverse field behavior of the ground-state splittings at resonances $k = 1$ and $k = 2$. Note that the splitting of the resonances allowed by symmetry, $k = 0$ and $k = 3$ [30], are mainly generated by the molecular anisotropy, showing weak variation for moderate transverse field values, in contrast to the strong variation shown by the forbidden resonances, $k = 1$ and $k = 2$, for which the transverse field is the only source of level mixing. The effect of the JT axis tilting is particularly significant in the case of resonance $k = 2$ (observed at the lowest temperature), for which a splitting magnitude on the order of $10^{-5}$ K is achieved for fields below ~250 G, while the ground-state splitting at resonance $k = 1$ remains more than an order of magnitude smaller for the same range of transverse field values. As shown above, intermolecular dipolar interactions can provide magnetic fields (~250 G) which are strong enough to induce a tunnel splitting in the $k = 2$ resonance of the level observed in the experiment. Meanwhile, their effect on the $k = 1$ resonance is nearly two orders of magnitude weaker, thereby explaining the absence of this QTM step in our studies of carefully aligned crystals (see Fig. 2).

The present results exemplify the remarkable influence of the molecular symmetry on the magnetic relaxation of molecular nanomagnets and provide the first pristine demonstration of spin selection rules on the QTM for a SMM. The observed behavior must be attributed to the extremely high crystalline quality of these complexes, enabling deep insights into fundamental quantum behavior that were previously impossible. Of special significance is the remarkable finding that a rotation of the ZFS tensors of the individual ions (in a manner consistent with the crystallographic symmetry) has a profound effect on the transverse field behavior of the tunnel splitting in resonances forbidden by the molecular symmetry. We have shown that a small Jahn-Teller axis tilt ($8.5^{\circ}$) is sufficient to increase the tunnel splitting value for the $k = 2$ resonance up to an observable level for transverse field magnitudes commonly provided by intermolecular dipolar interactions or as a result of weak disorder. Note that the Jahn-Teller axes of individual ions are almost never parallel with respect to each other in *real* structures and, according to our results, this combined with dipolar fields and/or disorder may be the ultimate reason behind the apparent absence of spin selection rules in previous studies of QTM in SMMs.

**ACKNOWDLEDGEMENTS:**

We gratefully acknowledge fruitful discussions with Eduardo Mucciolo. This work has been supported by the US National Science Foundation (DMR0737802, DMR0747587, DMR0506946, DMR0804408, DMR0239481 and CHE0714488), and by NSERC, PITP, and CIFAR in Canada.

absence of a transverse field, in order to initialize all measurements from the same magnetic configuration. Thermal avalanches (extremely frequent in this sample) restricted to a few points our results for resonance $k = 2$, which were obtained by sweeping the magnetic field at different angles with respect to the easy anisotropy axis of the molecules.

29. The observations in Fig. 4 are essentially insensitive to the orientation of the transverse field within the hard anisotropy ($x$-$y$) plane over the range of parameter space explored in this investigation. This agrees with the absence of any modulation in both the *dM*/*dH* peak magnitudes and EPR absorption peak positions (to within the resolution range of the techniques).

30. The tunnel splitting of resonance $k = 3$ also vanishes at zero transverse field when the Jahn-Teller axes of the Mn$^{III}$ ions are parallel to the $z$-axis (thin blue line in Fig. 4). In this situation, the $120^\circ$ rotations of the single-ion hard/medium axes generate a molecular 6th-order transverse interaction (i.e. $S_+^6 + S_-^6$), restricting QTM to resonances with $|\Delta m_S| = 6n$. It is the tilt, $\alpha$, of the Jahn-Teller axes away from the $z$-axis that imposes the $C_3$ rotational symmetry of the molecule, lowering the order of the transverse interaction (i.e. $[S_z,(S_+^3 + S_-^3)]$), and allowing QTM resonances with $|\Delta m_S| = 3n$ (thick blue line in Fig. 4).

**FIGURES (for review purposes):**

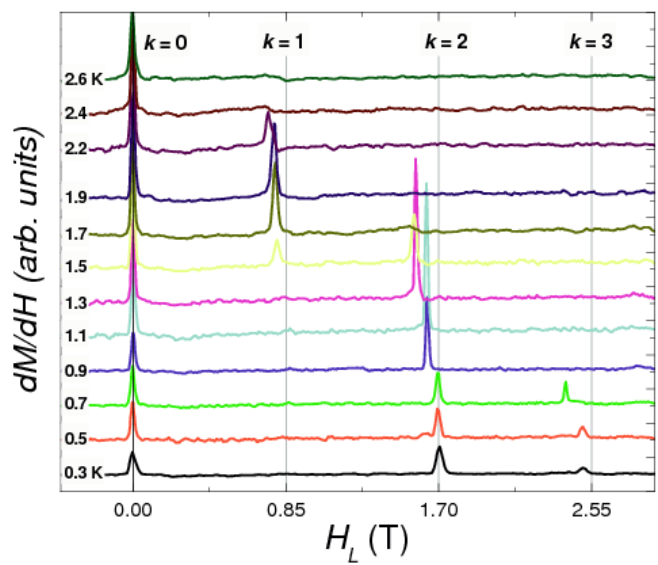

**FIG. 1:** (Color online) Field derivative of the magnetization curves obtained for a $Mn_3$-Cl single crystal at different temperatures, with the field swept at 1 T/min along the easy axes of the molecules.

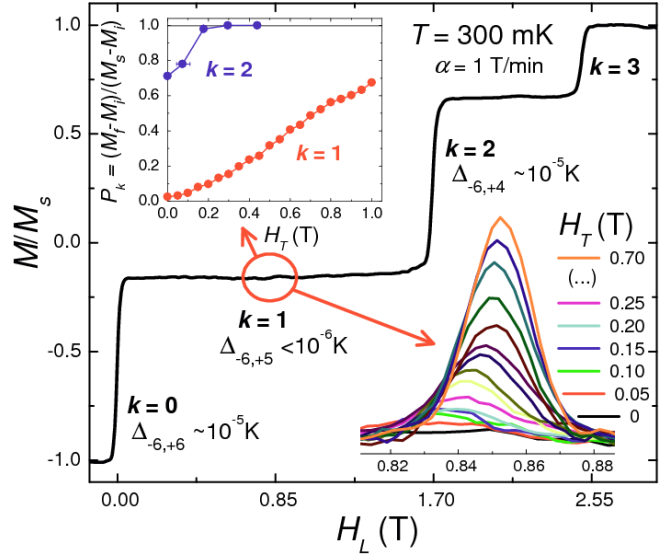

**FIG. 2:** (Color online) Magnetization versus longitudinal magnetic field recorded at 0.3 K in a $Mn_3$-Cl single crystal. The indicated tunnel splitting values are order of magnitude estimates from the change of magnetization at each resonance [27]. The insets show the increase of the QTM probability at resonances $k = 1$ and 2 (upper-left) and the dM/dH peak at resonance $k = 1$ (lower-right) upon application of a transverse magnetic field.

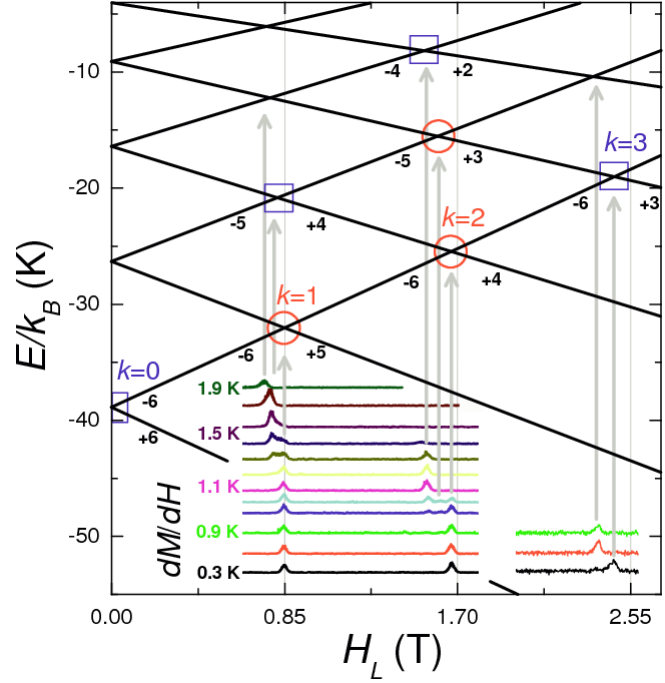

**FIG. 3:** (Color online) Energies of the Mn$_3$ $m_S$ levels as a function of the longitudinal magnetic field. Symbols highlight level crossings between different spin states, indicating the degeneracies (red circles) and avoided crossings (blue squares) expected for the $C_3$ symmetry of the structures. Inset: field derivatives of the magnetization curves ($dM/dH_L$) measured for a Mn$_3$-Br single crystal at different temperatures. Low field sweep rates, 0.05 T/min (data below 1.7 T) and 0.2 T/min (data above 1.7 T), where used to clearly observe and follow the evolution of resonance $k = 1$, which is absent for high sweep rates.

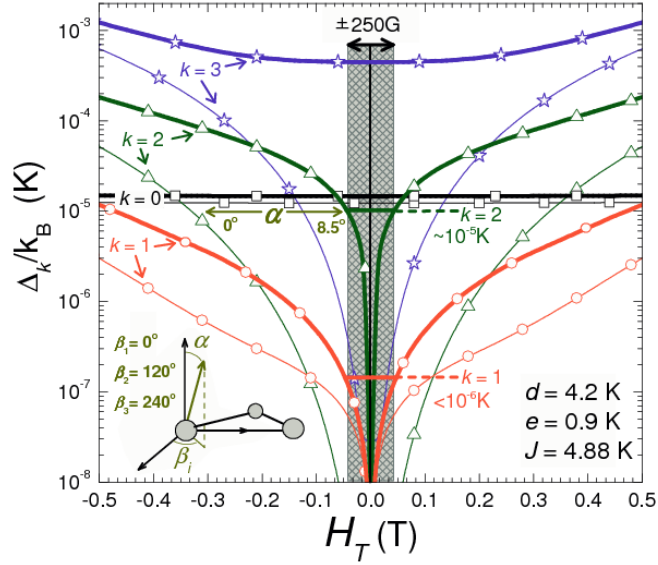

**FIG. 4:** (Color on-line) Ground state tunnel splittings associated with resonances $k = 0$ (black squares), $k = 1$ (red circles), $k = 2$ (green triangles) and $k = 3$ (blue stars) as a function of the transverse field, $H_T$, with the Jahn-Teller axes aligned along the $z$-axis (thin lines) and tilted $\alpha = 8.5°$ away from the $z$-axis (thick lines). A few symbols per curve have been added to help in their identification. The grey shaded region in the vicinity of zero transverse field represents the strength of the dipolar magnetic field (~250 G) felt by the Mn$_3$ molecules within the single crystal. The horizontal lines within the grey region indicate the order of magnitude of the splitting values attained at resonances $k = 1$ and $k = 2$ for a transverse field of 250 G.